\documentclass[draftclsnofoot, 12pt, onecolumn]{IEEEtran}
\pdfoutput=1
\usepackage{graphicx}
\usepackage{color}
\usepackage{placeins}
\usepackage{float}
\usepackage{tabularx,colortbl}
\usepackage{amsmath}
\usepackage{multirow}
\usepackage{hyperref}

\begin{document}
%
% paper title
% can use linebreaks \\ within to get better formatting as desired
\title{Full-Duplex Systems Using Multi-Reconfigurable Antennas}

% author names and affiliations
% use a multiple column layout for up to three different
% affiliations
\author{Elsayed Ahmed, Ahmed M. Eltawil, Zhouyuan Li, and Bedri A. Cetiner \footnote{Elsayed Ahmed and Ahmed M. Eltawil are with the Department of Electrical Engineering and Computer Science at the University of California, Irvine, CA, USA (e-mail: \{ahmede, aeltawil\}@uci.edu). Zhouyuan Li and Bedri A. Cetiner are with the Department of Electrical and Computer Engineering at Utah State University, Logan, Utah, USA (e-mail: zhouyuan.li@aggiemail.usu.edu, bedri.cetiner@usu.edu).}
}

% make the title area
\maketitle

\vspace{-0.5in}

\begin{abstract}
Full-duplex systems are expected to achieve 100\% rate improvement over half-duplex systems if the self-interference signal can be significantly mitigated. In this paper, we propose the first full-duplex system utilizing Multi-Reconfigurable Antenna (MRA) with $\sim$90\% rate improvement compared to half-duplex systems. MRA is a dynamically reconfigurable antenna structure, that is capable of changing its properties according to certain input configurations. A comprehensive experimental analysis is conducted to characterize the system performance in typical indoor environments. The experiments are performed using a fabricated MRA that has 4096 configurable radiation patterns. The achieved MRA-based passive self-interference suppression is investigated, with detailed analysis for the MRA training overhead. In addition, a heuristic-based approach is proposed to reduce the MRA training overhead. The results show that at 1\% training overhead, a total of 95dB self-interference cancellation is achieved in typical indoor environments. The 95dB self-interference cancellation is experimentally shown to be sufficient for 90\% full-duplex rate improvement compared to half-duplex systems.
\end{abstract}

%\begin{IEEEkeywords}
%Full-duplex systems, reconfigurable antennas, passive self-interference suppression.
%\end{IEEEkeywords}

% For peer review papers, you can put extra information on the cover
% page as needed:
% \ifCLASSOPTIONpeerreview
% \begin{center} \bfseries EDICS Category: 3-BBND \end{center}
% \fi
%
% For peerreview papers, this IEEEtran command inserts a page break and
% creates the second title. It will be ignored for other modes.
\IEEEpeerreviewmaketitle

\section{Introduction}
Due to the tremendous increase in wireless data traffic, one of the major challenges for future wireless systems is the utilization of the available spectrum to achieve better data rates over limited spectrum. Recently, full-duplex transmission, where bidirectional communication is carried out over the same temporal and spectral resources, was introduced as a promising mechanism that could potentially double the spectral efficiency of wireless systems. The main limitation impacting full-duplex transmission is managing the strong self-interference signal imposed by the transmit antenna, on the receive antenna, within the same transceiver. Recently, several publications~\cite{Ref1}-\cite{Ref14} have considered the problem of self-interference cancellation in full-duplex systems by investigating different self-interference cancellation techniques to mitigate the self-interference signal.

Self-interference cancellation techniques are divided into two main categories: passive suppression, and active cancellation. In passive suppression~\cite{Ref10}-\cite{Ref14}, the self-interference signal is suppressed in the propagation domain before it is processed by the receiver circuitry. In active cancellation techniques~\cite{Ref5}-\cite{Ref9}, the self-interference signal is mitigated by subtracting a processed copy of the transmitted signal from the received signal. Several experimental and analytical results show that the mitigation capability of active cancellation techniques is very limited, mainly due to the transmitter and receiver radio circuits' impairments~\cite{Ref15}-\cite{Ref18}. On the other hand, because it mitigates the signal in the propagation domain, passive suppression techniques mitigates both the self-interference signal and the transmitter noise associated with it. In addition, mitigating the self-interference signal in the propagation domain decreases the effect of the receiver noise and increases the dynamic range allocated for the desired signal. 

In this work, we propose a complete full-duplex system utilizing Multi-Reconfigurable Antenna (MRA). Per the authors' knowledge, this is the first reported full-duplex system utilizing MRA's with an experimentally proven 90\% rate improvement over half-duplex systems. MRA is a dynamically reconfigurable antenna that is capable of changing its proprieties (e.g. radiation pattern, polarization, and operating frequency) according to certain input configurations. The system performance is experimentally characterized in typical indoor environments using a fabricated MRA with 4096 dynamically configurable radiation patterns.
\subsection{Contribution}
The main contributions of the paper are as follows: first, we introduce the design and operating mechanism of a 2.5GHz MRA antenna. The MRA has 4096 possible modes of operation by configuring the surface geometry of the parasitic layer, where the 3x3 electrically small square-shaped metallic pixels (parasitic pixel surface) are connected by 12 PIN diode switches with ON/OFF status. The advantages of the proposed MRA design compared to the antenna presented in~\cite{Ref3Tony} can be summarized as follows: 1) The parasitic surface only consists of 3x3 metallic square pixels instead of 4x4 metallic rectangular pixels, thus the complexity of the MRA is reduced. 2) Real PIN diode switches have been used instead of ideal perfect open/short connection.

Second, a pattern selection mechanism to select the optimum pattern among the various MRA patterns is presented. Since the MRA has many radiation patterns, one can select the pattern that minimizes the received self-interference power. However, this method can not guarantee the optimal overall system performance, mainly because the selected pattern also affects the received signal-of-interest (the desired signal) power. To guarantee the best overall system performance, we developed a pattern selection mechanism that maximizes the received Signal-of-interest to Interferer Ratio (SIR) at the receiver input. Using MRA as a receive antenna in a full-duplex systems, the performance of the MRA-based passive self-interference suppression is experimentally investigated. The results show that, the MRA can achieve an average of 65dB of passive self-interference suppression, with a 45dB SIR gain compared to the case when an omni-directional antenna is used.

Third, since the MRA has to be trained in order to select the optimal pattern, training time, and training overhead are important design parameters that have to be investigated. In this paper, we present a detailed experimental analysis for the required MRA training time and training overhead in different indoor environmental conditions. In addition, a heuristic-based approach is proposed to reduce the training overhead by selecting a small suboptimal set of patterns among all MRA patterns. The results show that using the proposed heuristic, at 1\% training overhead with a suboptimal set of 300 patterns, 62dBs of passive suppression can be achieved with only a 3dB performance loss as compared to the optimal case.

Finally, a complete full-duplex system with a combined MRA-based passive suppression and conventional active self-interference cancellation is presented. The overall system performance is evaluated in different indoor environmental conditions. The results show that at 1\% training overhead, a total of 95dB self-interference cancellation is achieved in typical indoor environments. The 95dB self-interference cancellation is experimentally shown to be sufficient for 90\% full-duplex rate improvement compared to half-duplex systems at 5dBm transmit power.
\subsection{Prior Work}
Throughout the literature, passive self-interference suppression is achieved through one or a combination of the following four methods: (i) antenna separation, (ii) antenna isolation, (iii) antenna directionality, and (iv) antenna polarization. The applicability of each one of these methods depends on the application, and the physical constraints of the system. For example, in mobile applications with small device dimensions, the passive suppression achieved using antenna separation and isolation is very limited. However, in others systems (e.g. relay systems) where the transmit and receive antennas are not necessary collocated, antenna separation and isolation could achieve significant passive suppression. For instance, in~\cite{Ref20}-\cite{Ref21}, the use of a single pattern directional antenna and 4$-$6 m of antenna separation achieves $\sim$85dB of passive suppression. While in~\cite{Ref22}, using 5 m of antenna separation in addition to antenna isolation achieves 70dB of passive suppression. This large antenna separation might be acceptable in relay systems, but it is not acceptable in practical mobile applications. A more practical passive self-interference suppression method with relatively small antenna separation (e.g 20$-$40 cm) was introduced in~\cite{Ref6,Ref14}. The results show a maximum of 60dB passive suppression at 40cm antenna separation with cross polarization, and a metal shield between the antennas.

Recently, a comprehensive study for the achieved passive suppression using different combinations of the previously mentioned methods was introduced in~\cite{Ref13}. In~\cite{Ref13}, the passive suppression performance is characterized using two single-pattern directional antennas placed at different orientations, with different antenna separations ranging from 35-50 cm. The results show that in non reflective environment (e.g. Anechoic Chamber), a maximum of 72dB passive suppression could be achieved when absorptive shielding is present between the two antennas. While in a reflective room the maximum achievable passive suppression is reduced to 45dB due to the self-interference signal reflections.    

In contrast with the prior work, we focus on the deployment of full-duplex transmission in mobile indoor applications where the allowed antenna separation is very limited. Our approach could achieve 65dB of passive suppression at only 10 cm antenna separation in a reflective indoor environment, without any antenna shielding. Moreover, the directional antenna used in all prior work is a single pattern directional antenna. The lack of beam steering capability in such antennas might affect the signal-of-interest power in certain scenarios (e.g. when the desired signal is coming from the opposite direction of the antenna). On the other hand, the re-configurability feature in the full-duplex systems utilizing MRA attempts to maximize the SIR for any given scenario.     
\subsection{Paper Organization}
The rest of the paper is organized as follows; the MRA design and characteristics are presented in section II. In section III, the experimental framework and the experimental environment are described in details. Section IV presents the experimental results and discussions. In section V, the overall full-duplex system performance is introduced. Finally, section VI presents the conclusion.

\section{ANTENNA STRUCTURE AND WORKING MECHANISM}
\subsection{MRA Structure}
\begin{figure}[!t]
\begin{center}
\noindent
  \includegraphics[width=6.5in, height=3in]{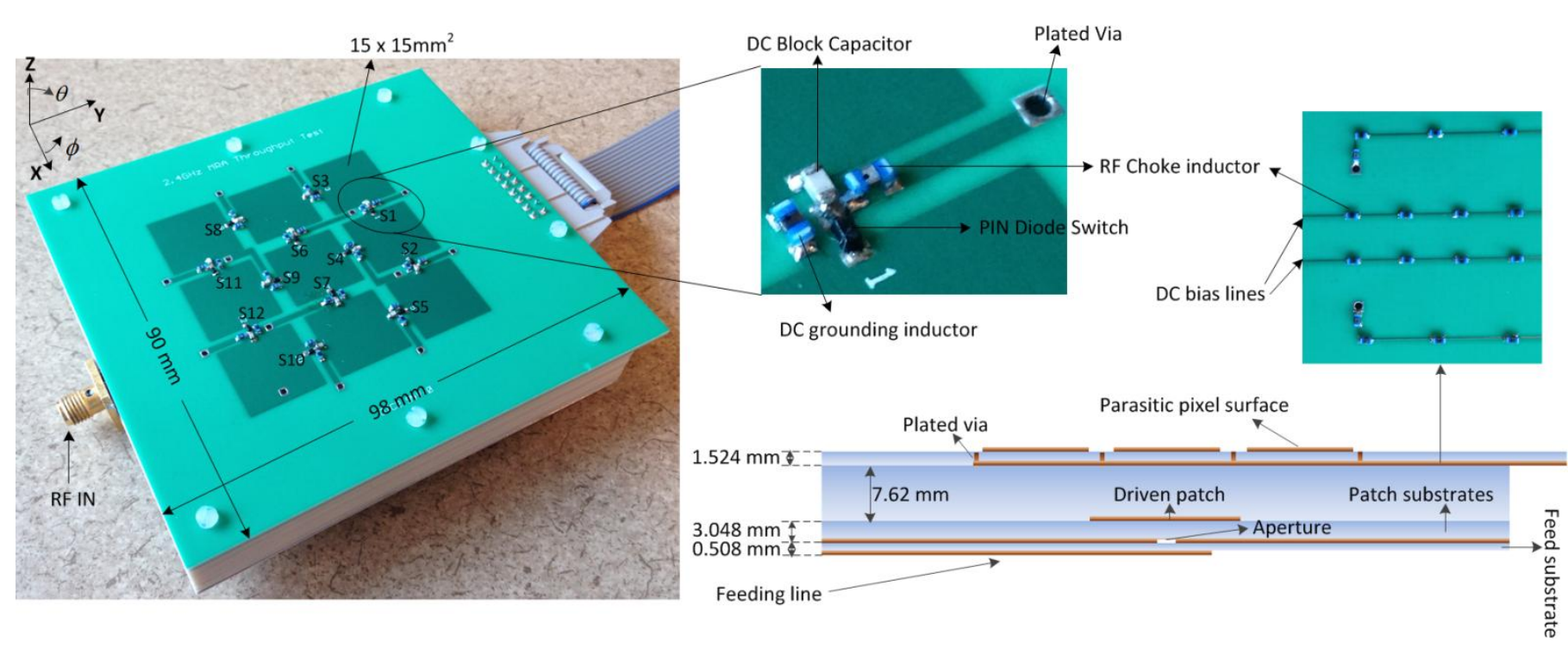}
  \caption{3-D schematic and the cross section view of the MRA.\label{Fig1TonyLabel}}
\end{center}
\end{figure}

The 3-D schematic and cross section view of the MRA are depicted in figure~\ref{Fig1TonyLabel}. This MRA employed an aperture-coupled feed mechanism for RF feeding similar to the MRA presented in~\cite{Ref3Tony}. The main two components of the MRA architecture are, namely, the driven patch antenna and parasitic layer. The driven patch (19.3x19.3 mm$^2$) is designed to operate in the frequency band of 2.4-2.5 GHz and fed by a 50-Ohm microstrip line through an aperture (21.4x1.4 mm$^2$) etched on the center of the common ground plane. The feed layer (90x90x0.508 mm$^3$) and patch layer (90x90x3.048 mm$^3$) are built respectively by using the substrate Rogers 4003C ($\epsilon_r$ = 3.55, $tan\  \delta$ = 0.0027)~\cite{Ref4Tony}. The same substrate (98x90x1.524 mm$^3$) is used to form the parasitic layer above the driven patch. Notice that there is a 7.62 mm gap between parasitic layer and driven patch antenna, where the gap is filled with the RO4003C.

The reconfigurable parasitic surface, which consists of 3x3 square-shaped metallic pixels connected by 12 PIN diode switches with ON/OFF status, is formed on the top surface of the parasitic layer with individual pixel size being 15x15 mm$^2$. Thus the geometry of the parasitic surface can be configured by switching ON/OFF the 12 PIN diode switches, which are marked as S1-S12 in figure~\ref{Fig1TonyLabel}. DC bias lines for controlling the PIN diode switches are also formed on the parasitic layer but on the backside of the substrate. Vias are plated through the parasitic layer so that DC bias lines can be connected to the PIN diode switches on the parasitic surface.

Four different kinds of lumped components are used on the parasitic layer as shown in figure~\ref{Fig1TonyLabel}: 1) PIN diode switches are used in between all rectangular pixels. Metallic pixels are connected/disconnected by switching ON/OFF the PIN diode switches to change the geometry of the parasitic surface, which in turn change the current distribution, and thus RF characteristic. 2) Inductors are placed along the DC bias lines as RF chokes. The SRF (self resonant frequency) of the RF choke is chosen to be around 2.5 GHz, thus RF chokes would appear as high impedance in the ISM band to minimize the current on the bias lines, thus minimizing the effect of the bias lines on the antenna performance. 3) Inductors are also placed in between all the rectangular pixels to connect all the pixels together. In this manner, all the pixels can be DC grounded together to provide Ground for DC biasing purpose. The SRF of these inductors was chosen to be the same value as RF chokes to keep the high RF impedance between pixels.  4) DC block capacitors are used to properly bias the PIN diode switches as shown in figure~\ref{Fig1TonyLabel}. The SRF of DC block capacitor is around 2.5 GHz to provide low RF impedance in the ISM band. In this way, the effect of DC block capacitor on RF performance is minimized.

\subsection{Working Mechanism}
\begin{figure}[!t]
\begin{center}
\noindent
  \includegraphics[width=5in]{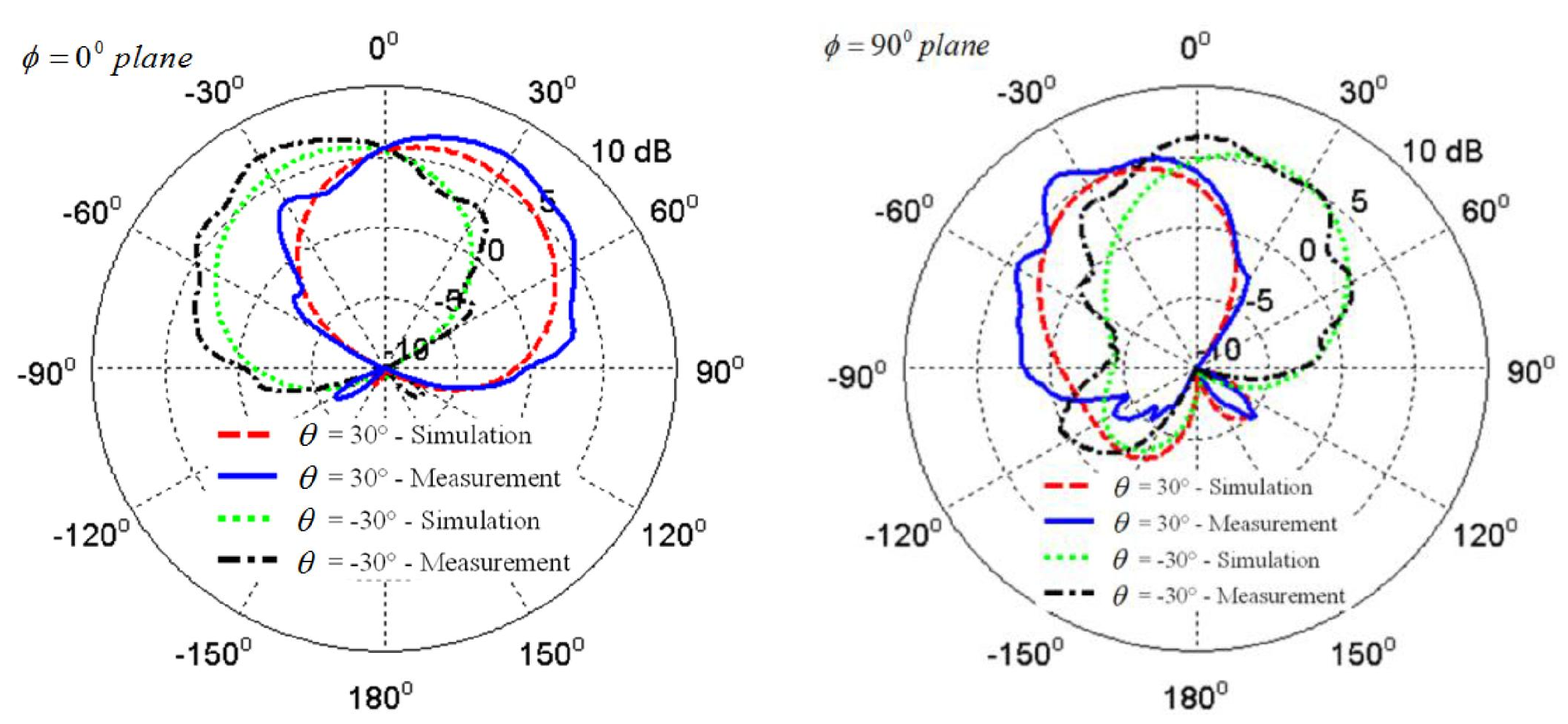}
  \caption{Simulated and measured radiation patterns for four different MRA modes at 2.45 GHz.\label{Fig5TonyLabel}}
\end{center}
\end{figure}

The working mechanism of the antenna system, which is composed of one driven antenna and multiple parasitic elements, can be described by the theory of reactively controlled directive arrays developed by R. F. Harrington~\cite{Ref5Tony}. It was shown that the main beam direction of the driven antenna can be directed into a desired direction by the proper reactive loading of the parasitic elements. In the presented MRA, the proper reactive loading corresponds to a specific geometry of the parasitic pixel surface, which is obtained by switching ON/OFF the PIN diode switches between adjacent pixels of this surface. Switching ON and OFF the PIN diode switches placed on the MRA surface creates 4096 different modes of operation each with unique MRA radiation pattern. As an example, figure~\ref{Fig5TonyLabel} shows the simulated and measured MRA radiation pattern for four different modes of operation, showing good agreement between the simulated and measured patterns.

%An example of the beam steering capability of the MRA is shown in figure~\ref{Fig4TonyLabel}a. It is possible to steer the main beam of the MRA in multiple directions in the semi-sphere space. The presented MRA has a total of 4096 different radiation patterns spanning the whole semi-sphere space. figure~\ref{Fig4TonyLabel}b shows an example for one of the 4096 radiation patterns.
%
%\begin{figure}[!ht]
%\begin{center}
%\noindent
%  \includegraphics[width=4in]{figure4Tony.pdf}
%  \caption{The beam-steering example of the MRA: $\theta \in \left\{-30^0,0^0,30^0\right\}; \phi \in \left\{0^0,45^0,90^0,135^0\right\}$.\label{Fig4TonyLabel}}
%\end{center}
%\end{figure}
%

\section{Experimental Framework and Environment}
Due to the significant dependence of the full-duplex system performance on hardware impairments and the surrounding environments, experimental analysis is extremely important for performance characterization in full-duplex systems. In addition to hardware impairments, the use of a directional antenna at such small antenna separation creates a near-field effect that is difficult to account for at every possible scenario. In this section, the experimental setup, framework, and experimental environment are described in details.
\subsection{Experimental Setup}
A complete full-duplex system is constructed using the Universal Software Radio Peripheral (USRP) software defined radio (SDR) platform~\cite{Ref23}. Each USRP contains a Radio Frequency (RF) transceiver and a Field Programmable Gate Array (FPGA). All USRP's are connected to a host PC through a Gigabit Ethernet connection. The baseband signal processing is performed over the host PC. The baseband signals are streamed to/from the USRPs at a rate of 25M sample/sec. The RF transceivers are then used for real time signal transmission and reception. All experiments are performed in the ISM band at 2.5Ghz carrier frequency with a 10Mhz signal bandwidth. All USRPs are synchronized to one reference clock.

\begin{figure}[!ht]
\begin{center}
\noindent
  \includegraphics[width=3in]{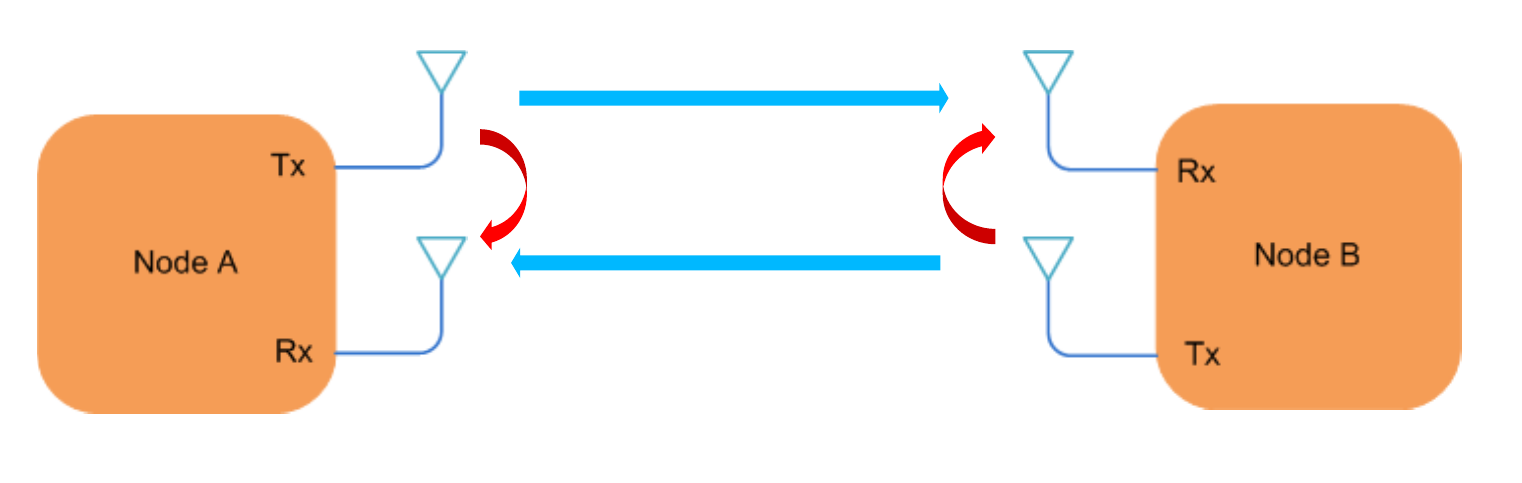}
  \caption{Two nodes full-duplex system.\label{Fig1Label}}
\end{center}
\end{figure}

\begin{figure}[!ht]
\begin{center}
\noindent
  \includegraphics[width=3in]{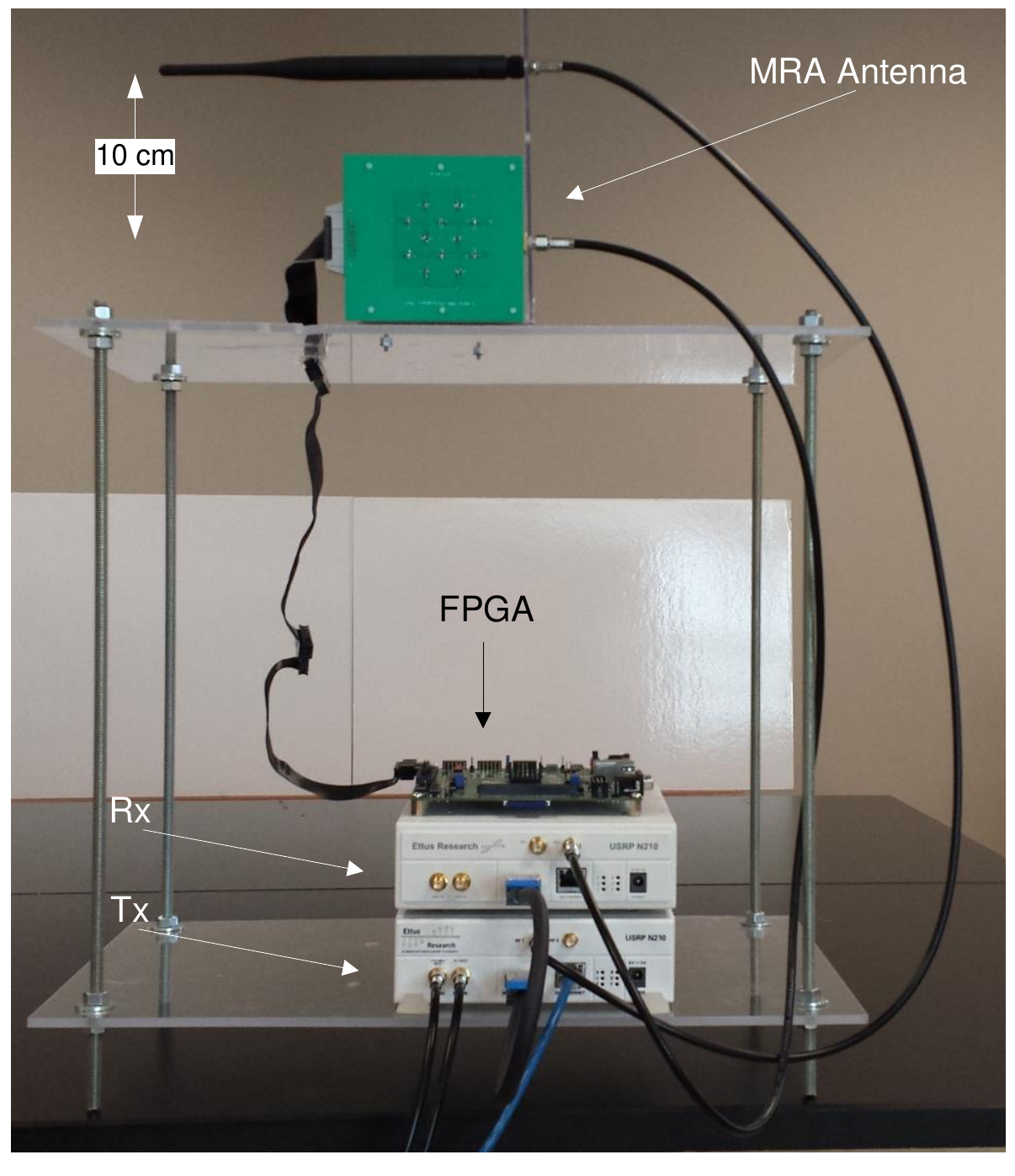}
  \caption{Full-duplex wireless node with MRA.\label{Fig2Label}}
\end{center}
\end{figure}

As shown in figure~\ref{Fig1Label}, the full-duplex system consists of two nodes communicating in a full-duplex manner. Each node is equipped with one transmit antenna and one receive antenna. In this paper, a dipole omni-directional antenna is used as transmit antenna, while the MRA is used as receive antenna. Both transmit and receive antennas have the same antenna polarization\footnote{There are many other antenna configurations that could be used. For example, the MRA could be use as transmit antenna, or both transmit and receive antennas. Furthermore, cross antenna polarization could be used.}. The MRA antenna has a total of 4096 different radiation patterns. The pattern selection is performed through a 12-lines digital control cable driven from an FPGA on a Zedboard~\cite{Ref24}. The timing of all USRPs and the FPGA that drives the antenna switches are aligned with one reference Pulse Per Second (PPS) signal. Figure~\ref{Fig2Label} shows a typical structure for a full-duplex node using MRA antenna. Another full-duplex system architecture where both transmit and receive antennas are omni-directional antennas is used for comparison purposes
\subsection{Experimental Framework}
\begin{figure}[!ht]
\begin{center}
\noindent
  \includegraphics[width=3in]{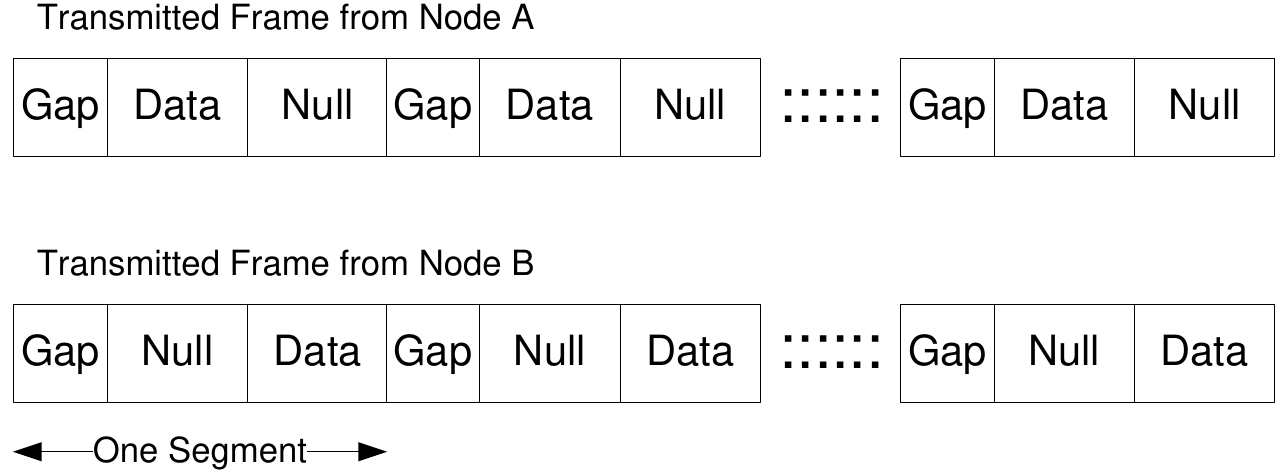}
  \caption{MRA training frame structure.\label{Fig3Label}}
\end{center}
\end{figure}

In this paper, two different frameworks are used for performance  characterization: the passive suppression characterization framework, and the complete system framework. In the passive suppression characterization framework, the full-duplex system is used to characterize the achieved passive self-interference suppression for each MRA radiation pattern at different environmental conditions. For measurement purposes, in this framework, the received SIR defined as the ratio between the received signal-of-interest power and the received self-interference power is used as a performance metric. The passive suppression characterization frame structure is shown in figure~\ref{Fig3Label}. Each transmission frame consists of $L$ segments, where $L$ is the number of antenna patterns that need to be characterized. Each segment contains three intervals: Gap interval, Data interval, and Null interval. The Data and Null intervals have the same length and are alternating between the two nodes. The MRA radiation pattern is changed at the segment edge. The Gap interval is used to account for the MRA switching time. During the Data interval, the node is transmitting a training sequence, while during the Null interval the node is silent. At the receiver side, the transmitted frames from each node are combined and received by the MRA antenna. In the combined frame, each segment will contain a self-interference portion and a signal-of-interest portion. The received signal strength is calculated for each portion to obtain an estimate for the received self-interference and signal-of-interest power.

The complete system framework is used to characterize the overall full-duplex system performance when the MRA-based passive self-interference suppression is combined with the conventional digital cancellation technique. In this framework, two different performance metrics are used: the overall self-interference cancellation, and the achievable full duplex rate. The transmission frame structure in the complete system framework consists of two main intervals: the MRA training interval, and the data transmission interval. During the MRA training interval, the MRA patterns are trained and the optimum pattern is selected. During the data transmission interval, the full-duplex data transmission takes place between the two communicating nodes. During MRA training interval, a frame structure similar to the one described in the passive suppression characterization framework is used. On the other hand, the data transmission interval consists of several data frames that have the same frame structure as in the 802.11n systems~\cite{Ref25}. Each frame consists of several Orthogonal Frequency Division Multiplexing (OFDM) symbols with 64 subcarriers in each symbol. At the beginning of each data frame, training symbols are transmitted for channel estimation purposes. After channel estimation, digital self-interference cancellation is performed to mitigate the residual self-interference signal.
\subsection{Practical Aspects}
Since the optimum pattern selection process involves extensive training, training time and training overhead are important parameters that have to be investigated. According to the MRA training frame structure, the training time and training overhead are a function of two main parameters: the number of MRA patterns that have to be trained, and the segment length. In this section, the required minimum segment duration is discussed (discussion related to the number of MRA patterns that have to be trained is presented in the Section IV). 

The segment duration is a function of the Gap and the Data intervals' length. The Gap interval lenegth is directly proportional to the MRA switching time which is function of the MRA switching circuitry. In the current design, the MRA switching time is $\sim$0.5us. The length of the Data interval depends on how the received signal strength is calculated. For example, if the received signal strength is calculated in the digital domain, the Analog to Digital Converter (ADC) sampling rate and the allowable timing offset will determine the minimum Data interval length. Based on our extensive experiments, approximately 30 time-domain samples are enough to obtain a good estimate for the received signal strength. Therefore, using 40Mhz ADC sampling rate, the required minimum segment duration is 2us (0.5us for antenna switching, and 1.5us for Data and Null intervals per segment). This time could be reduced to 1.25us if the ADC sampling rate is doubled to 80Mhz, which is a practical sampling rate in current wireless systems
\begin{figure}[!ht]
\begin{center}
\noindent
  \includegraphics[width=3in]{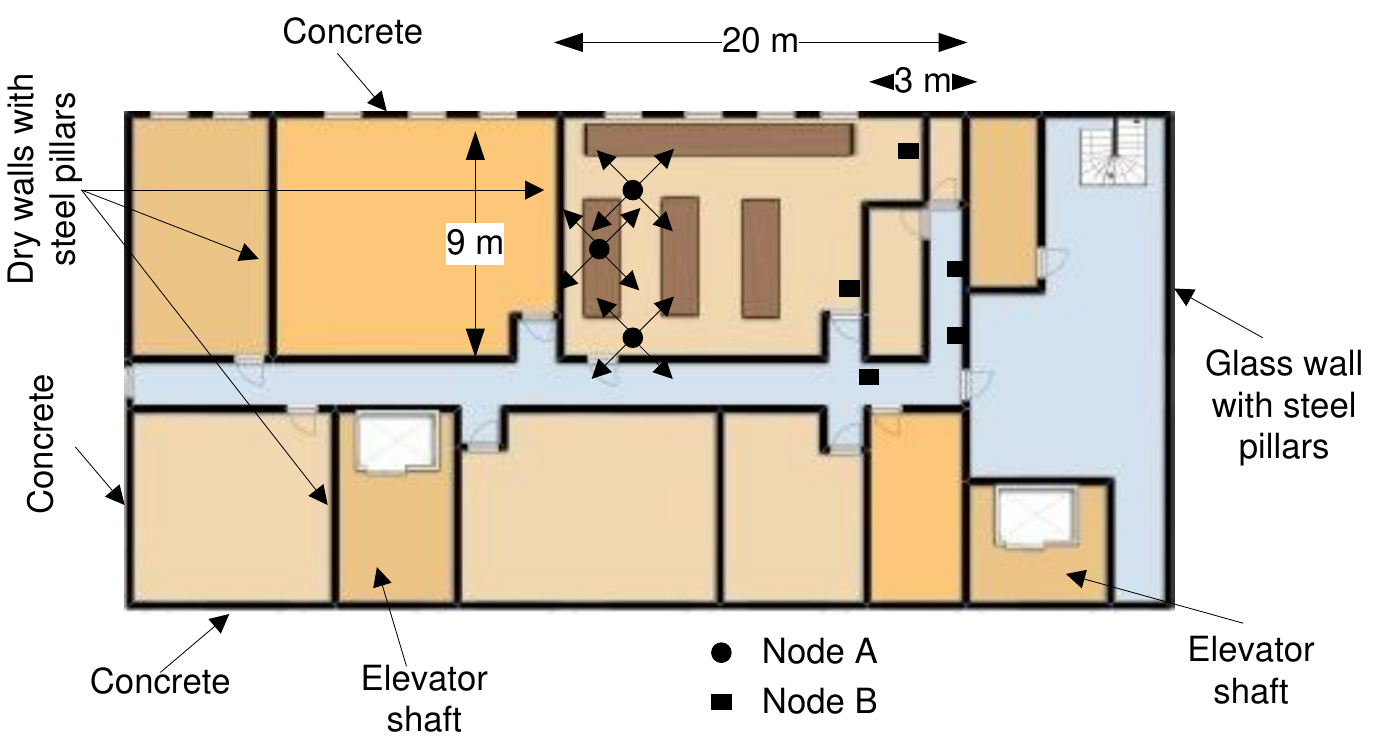}
  \caption{Floor plan for the area where the experiments are conducted.\label{Fig4Label}}
\end{center}
\end{figure}
\subsection{Experimental Environment}
The experimental analysis is conducted in the Wireless Systems and Circuits Laboratory (WSCL) within Engineering Hall at the University of California, Irvine. Figure~\ref{Fig4Label} shows a floor plan for the area where the experiments are performed, and presents a typical laboratory environment with measurement workstations, tables, metallic surfaces, etc. The outer walls of the building are either concrete walls or glass walls with steel pillars. While, the inner walls are dry walls with steel pillars.

To enrich the experimental analysis, the two communicating nodes are placed at different positions inside and outside the laboratory to create a variety of Line Of Sight (LOS) and non-LOS environments. In addition, different MRA orientations are tested such that the two communicating nodes are facing each other, opposite to each other, or side to side. To emulate typical conditions, the experiments are performed in both semi-static and dynamic environments. In a semi-static environment, the area is static with no moving personnel in the near area. While in dynamic environments, normal laboratory activities are maintained with moving personnel during the experiment time.

\section{Experimental Results}
In this section, the performance of the MRA-based passive suppression is characterized and discussed. The performance is also compared to the conventional omni-directional antenna based passive suppression. In addition, we present a heuristic-based approach to reduce the overall MRA training time by reducing the number of MRA patterns that need to be trained. The performance of the heuristic-based approach is compared to the optimal case where all MRA patterns are trained. Finally, the MRA training overhead and training periodicity are characterized and discussed.
\subsection{MRA-based Passive Self-interference Suppression}
In this part, the passive suppression characterization framework is used to characterize the achieved MRA-based passive Self-interference suppression. The performance is evaluated at different transmit power values ranging from $-$10dBm to 10dBm. Each run lasts for several seconds. In each run, all the 4096 MRA patterns are trained, and the pattern that maximizes the SIR is selected.

\begin{figure}[!ht]
\begin{center}
\noindent
  \includegraphics[width=3in]{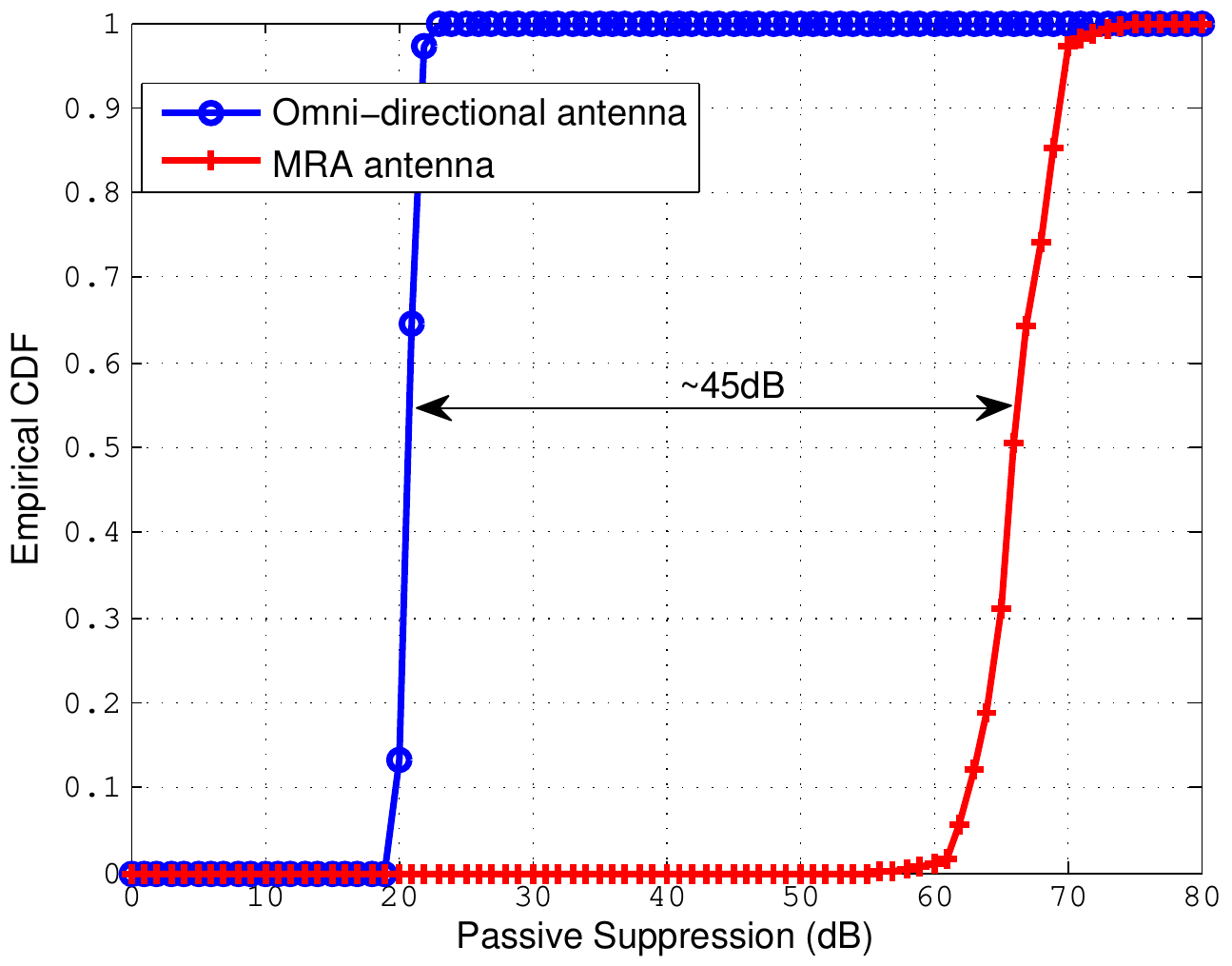}
  \caption{CDF of omni-directional antenna and MRA-based passive self-interference suppression.\label{Fig5Label}}
\end{center}
\end{figure}

Figure~\ref{Fig5Label} shows the empirical Cumulative Distribution Function (CDF) of the achieved passive self-interference suppression for both MRA and omni-directional antenna cases. The passive suppression is defined as the ratio between the transmit power and the received self-interference power at the antenna output. The CDF is calculated over time for all different runs and transmit power values. The results show that, using MRA achieves an average of 65dB passive suppression, with 45dB passive suppression gain compared to omni-directional antenna.

Since the selected MRA pattern affects the received signal-of-interest power, the achieved passive suppression amount is not sufficient to characterize the overall system performance. Instead, the effect of the MRA on the received signal-of-interest power should be also considered. The received signal-of-interest power is affected by both the MRA pattern, and the distance between the two communicating nodes. Therefore, to eliminate the distance factor and focus only on the MRA effect, the signal-of-interest power loss is used as a performance metric instead of the absolute value of the received signal-of-interest power. The signal-of-interest power loss is defined as the received signal-of-interest power ratio between the MRA case and the omni-directional antenna case for the same experimental environment. 

\begin{figure}[!ht]
\begin{center}
\noindent
  \includegraphics[width=3in]{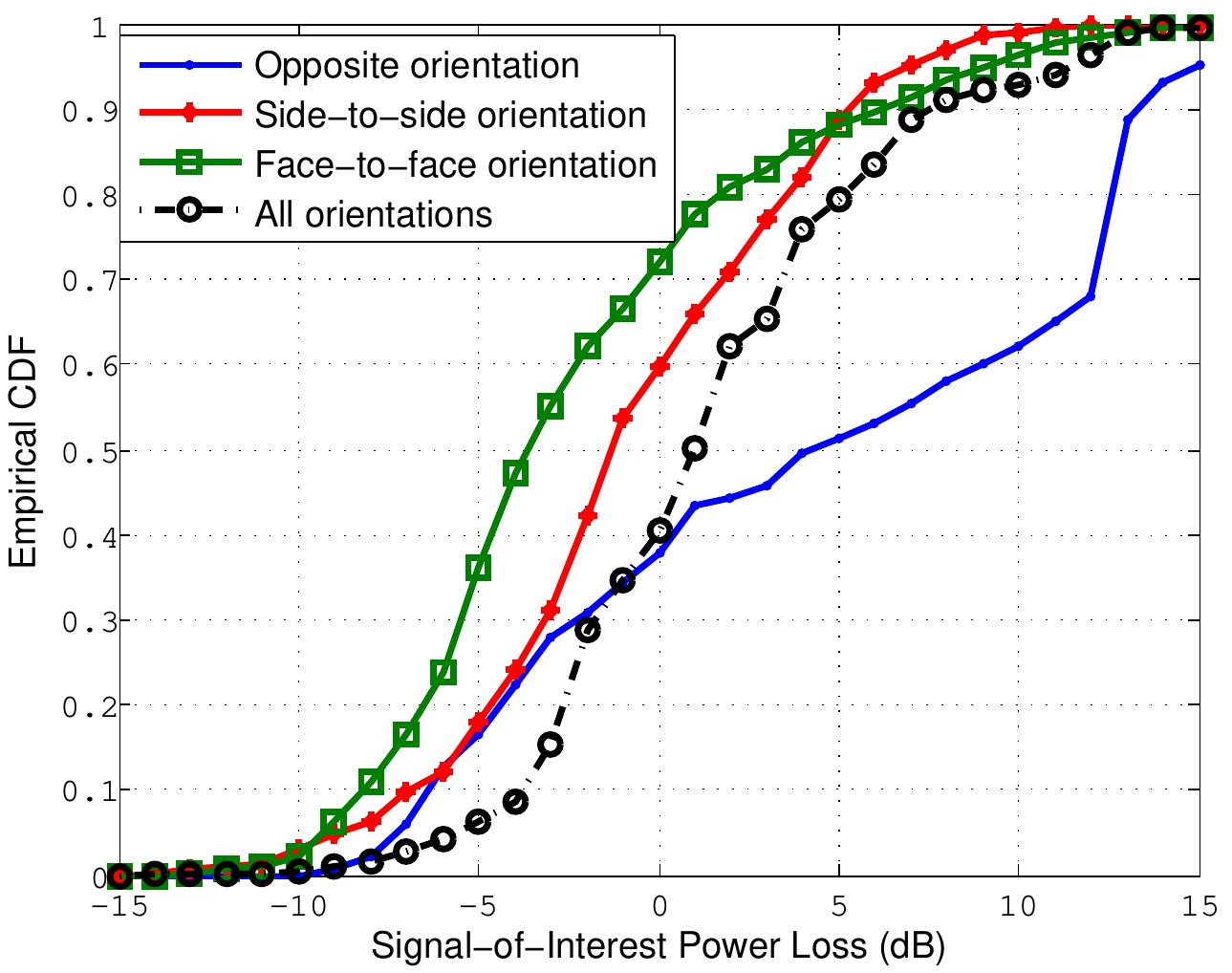}
  \caption{CDF of the Signal-of-interest power loss for different experimental environments.\label{Fig6Label}}
\end{center}
\end{figure}

Figure~\ref{Fig6Label} shows the empirical CDF of the signal-of-interest power loss for three different experimental environments, in addition to the average CDF for all environments. The description of the three different environments is as follows: in the opposite orientation environment, the MRA antennas in the two communicating nodes are placed back-to-back, such that the back side of the MRA at a node is facing the other node. The face-to-face orientation is the contrary of the opposite orientation. In the side-to-side orientation, the side of the MRA at one node is facing the other node. The main difference between the opposite orientation and the face-to-face orientation is that in the opposite orientation, the MRA is receiving most of the signal-of-interest power through its back loops which generally has small antenna gain. However, in the face-to-face orientation, most of the power is received through the main loops of the MRA which generally has high gain due to antenna directivity. Therefore, it is expected to have signal-of-interest power loss in the opposite orientations, while in the face-to-face orientation, the MRA is supposed to achieve signal-of-interest power gain. As shown in figure~\ref{Fig6Label}, an average of 5dB loss in the signal-of-interest power is expected in the opposite orientation environments, while an average signal-of-interest power gain of 4dB and 1dB is achieved in face-to-face and side-to-side orientations respectively. As an average over all different orientations, an average signal-of-interest power loss of 1dB is expected when the MRA is used. Compared to the 45dB self-interference suppression gain achieved by MRA, 1dB signal-of-interest power loss is negligible. 
\subsection{Suboptimal Pattern-set Selection Heuristic}
\begin{figure}[!ht]
\begin{center}
\noindent
  \includegraphics[width=3in]{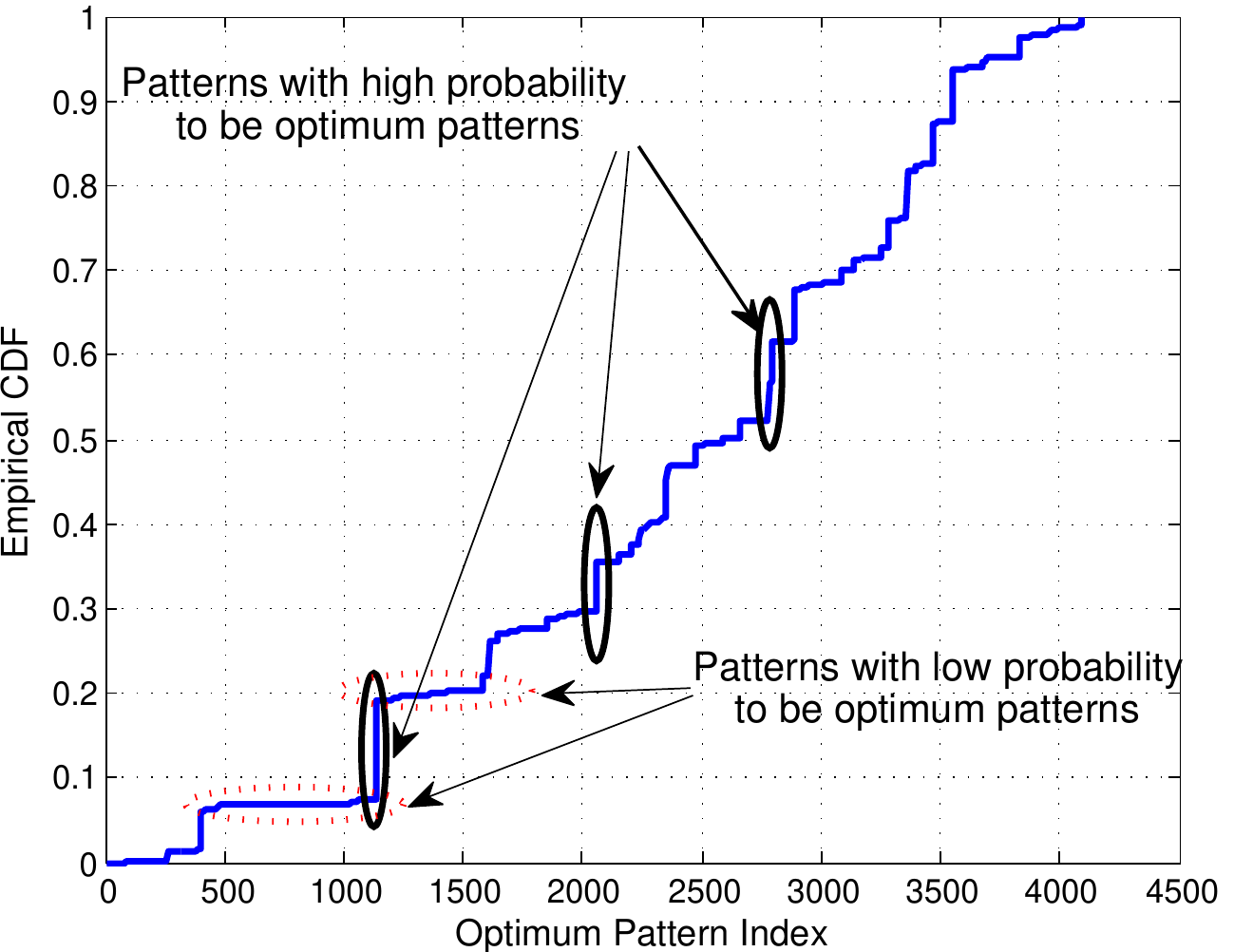}
  \caption{CDF of the index of the optimal MRA pattern.\label{Fig7Label}}
\end{center}
\end{figure}

While using the MRA as described leads to significant gains, the investment required to obtain 4096 modes is prohibitive. The goal of this section is to identify a heuristic that can reduce training overhead. To address this issue, we calculate the distribution of the optimal MRA pattern over time and for different environmental conditions. The calculated distribution is used to check if the optimal pattern index is localized or spans the whole range from 1 to 4096. The results in figure~\ref{Fig7Label} show that, the optimum pattern index spans the whole range, but it is not uniformly distributed. In fact, the results show that there are some patterns that have low or even zero probability to be among the optimum patterns, while other patterns have high probability to be among the optimum ones.

While one viable choice may be to exclude patterns with low probability of being optimal, it is important to take into account the degree of "sub-optimalty". In fact, for a pattern, to have a low (or zero) probability of being optimum does not necessary means that the pattern achieves poor performance. For instance, among those low probability patterns there are two categories: i) patterns that achieve good performance that is slightly less than the performance of the optimal pattern, and ii) patterns with poor performance that is significantly less than that of the optimal pattern. Although they have significant performance difference, the probability criterion does not differentiate between those two categories, because they are both considered non-optimal. Accordingly, a better selection criterion should involve the self-interference suppression performance for each pattern, not only the probability of being among the optimum patterns or not.

For further clarification, consider that in full-duplex systems, the self-interference signal arrives at the receive antenna in two main components: the LOS component through the direct link between the transmit and receive antennas, and the non-LOS component due to the reflections. Due to the close proximity of the transmit and receive antennas, the LOS component is much higher than the non-LOS component. Therefore any MRA pattern with high gain in the LOS direction most likely will achieve poor performance, thus this pattern has to be avoided. In fact, the optimal patterns are the patterns that are capable of suppressing not only the LOS component but also part of the non-LOS component.

\begin{figure}[!ht]
\begin{center}
\noindent
  \includegraphics[width=3in]{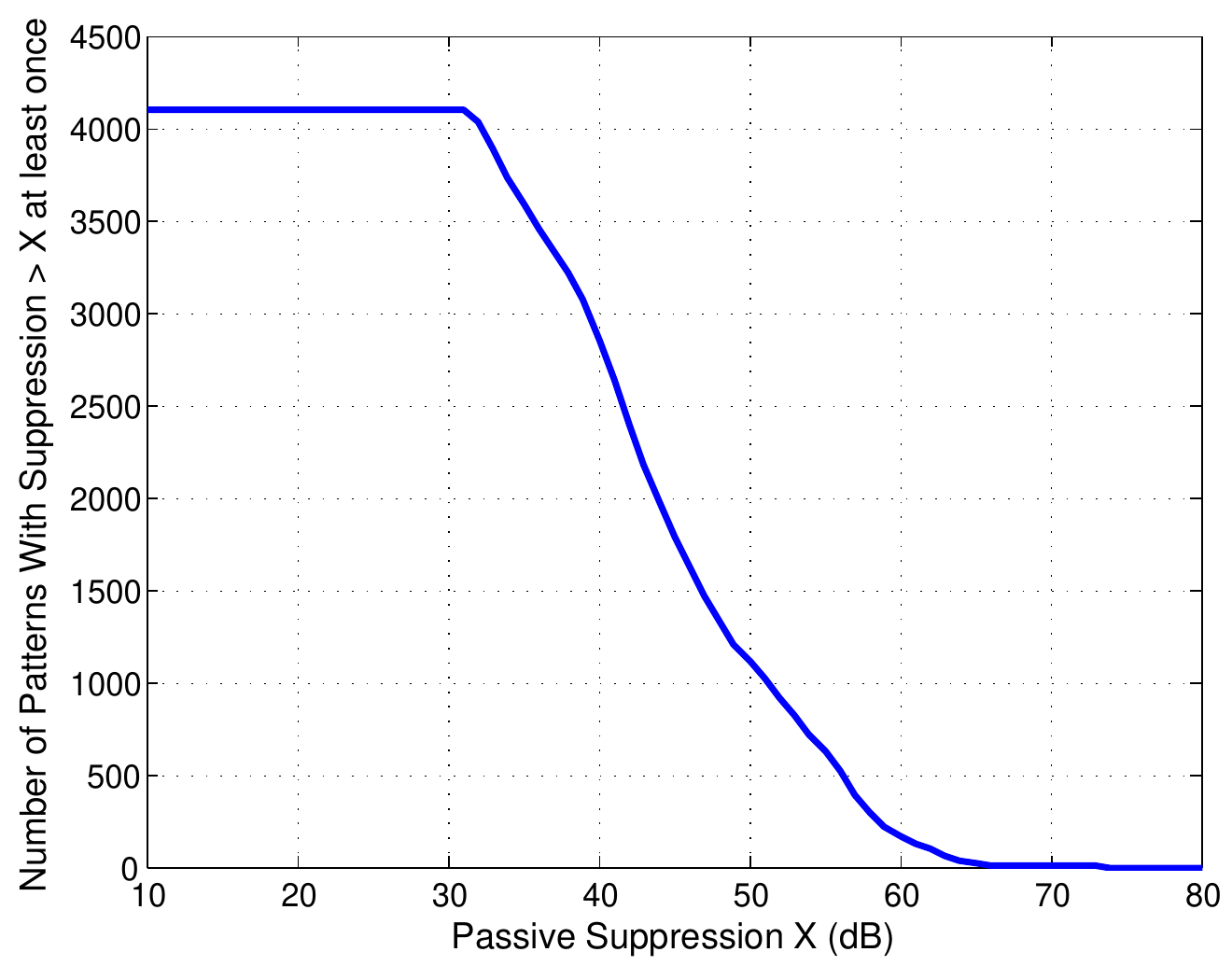}
  \caption{Number of MRA patterns that are capable of achieving passive suppression > $X$ in at least one of the tested scenarios.\label{Fig8Label}}
\end{center}
\end{figure}

Accordingly, based on the achieved self-interference suppression for each MRA pattern, we developed a heuristic-based approach to select a suboptimal set of patterns that are expected to achieve the best performance. First, we run the system in 16 different environments that includes a variety of LOS, non-LOS, semi-static, and dynamic scenarios each with 4 different orientations (opposite, face-to-face, and two side-to-side orientations). In each run, the achieved passive self-interference suppression for each one of the 4096 MRA patterns is calculated. We set a certain threshold $X$ that represents a desired passive self-interference suppression amount. Then, the patterns that achieves passive suppression $> X$ at any time in any environment are selected. Basically, we select the patterns that are capable of achieving passive suppression $> X$ at least once. Therefore, any pattern that is not selected is guaranteed to have passive suppression less than $X$ in all tested scenarios. The results in figure~\ref{Fig8Label} show the number of patterns that are capable of achieving passive suppression $> X$ at least once for different values of the threshold $X$. For instance, the results show that there are 1000, and 300 patterns capable of achieving passive suppression $>$ 52dB and 58dB respectively.

\begin{figure}[!ht]
\begin{center}
\noindent
  \includegraphics[width=6.5in]{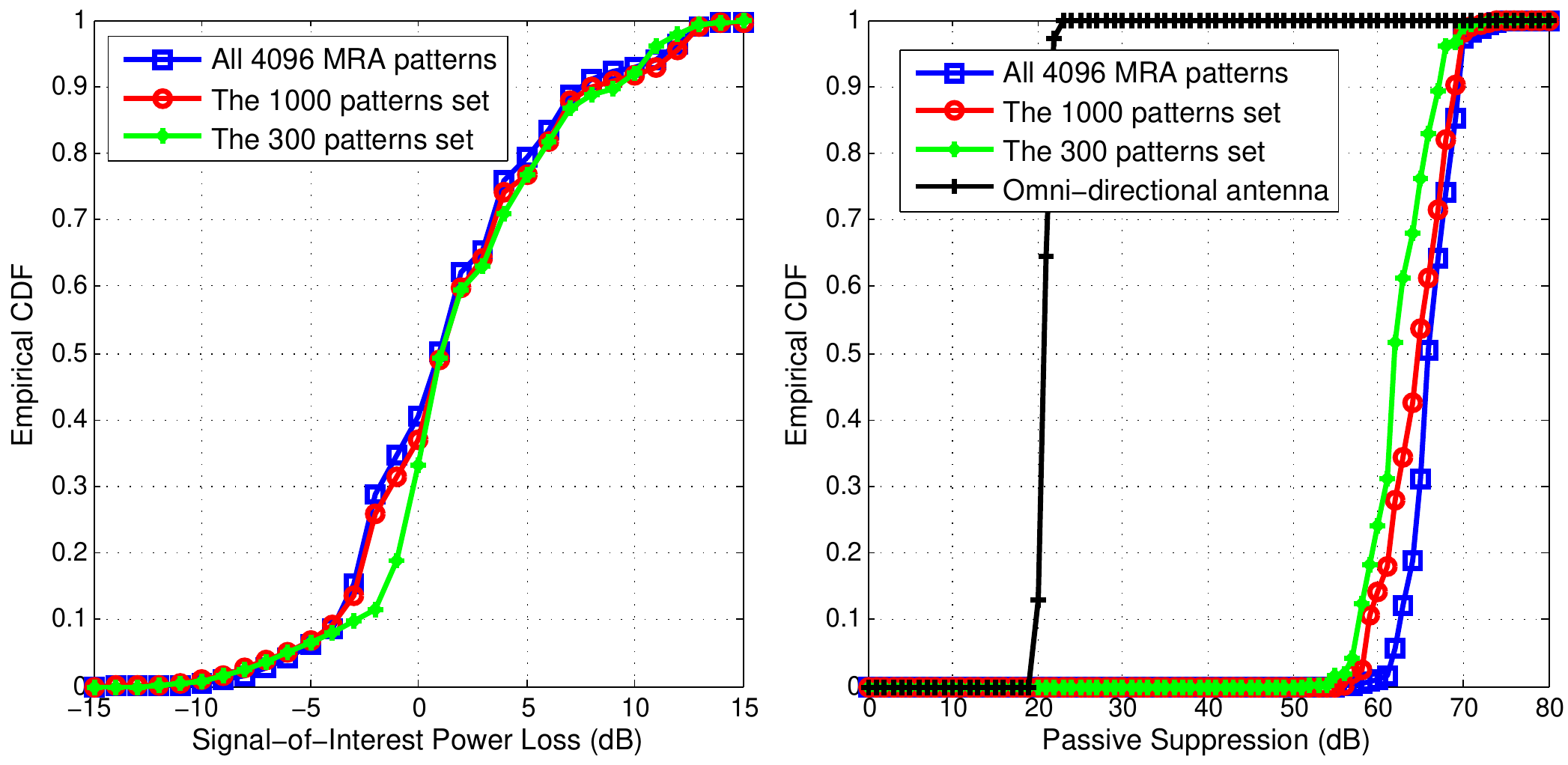}
  \caption{CDF of passive self-interference suppression and signal-of-interest power loss with different MRA pattern sets.\label{Fig9Label}}
\end{center}
\end{figure}

In order to test the accuracy of the proposed heuristic we selected two different suboptimal set of patterns with passive suppression threshold $X =$ 52dB and 58dB respectively. The first set contains 1000 patterns, and the second set contains 300 patterns. The performance of the selected sets is characterized in more than 20 different experimental environments that are different from the 16 environments used to select the suboptimal sets\footnote{The experimental environments in this/following analysis are different from the environments used to select the suboptimal sets in the sense that the positions of the two communicating nodes are changed and different orientations spanning the whole 360$^o$ are used.}. Figure~\ref{Fig9Label} shows the CDF of both passive self-interference suppression and signal-of-interest power loss for the selected sets as well as the optimal 4096-patterns set. The results show that the 300-patterns set achieves an average of 62dB passive self-interference suppression with 3dB loss compared to the optimum 4096 patterns set, but at $\sim$14 times less training time. Also, at $\sim$4 times less training time, the 1000-patterns set achieves an average of 64dB passive self-interference suppression. On the other hand, from signal-of-interest perspective, the results show that the 1000- and 300-patterns sets achieve almost the same performance as the optimal 4096-patterns set.
\subsection{MRA Training Overhead}
Due to its significant effect on the overall system capacity, training overhead is an important parameter that should be investigated. The training overhead is defined as the ratio between the training duration and the useful data duration. In the proposed full-duplex system, the training overhead is a function of two main parameters: the number of MRA patterns that need to be trained, and the re-training period. The re-training period is defined as the minimum time between two successive training intervals.

\begin{figure}[!t]
\begin{center}
\noindent
  \includegraphics[width=6.5in]{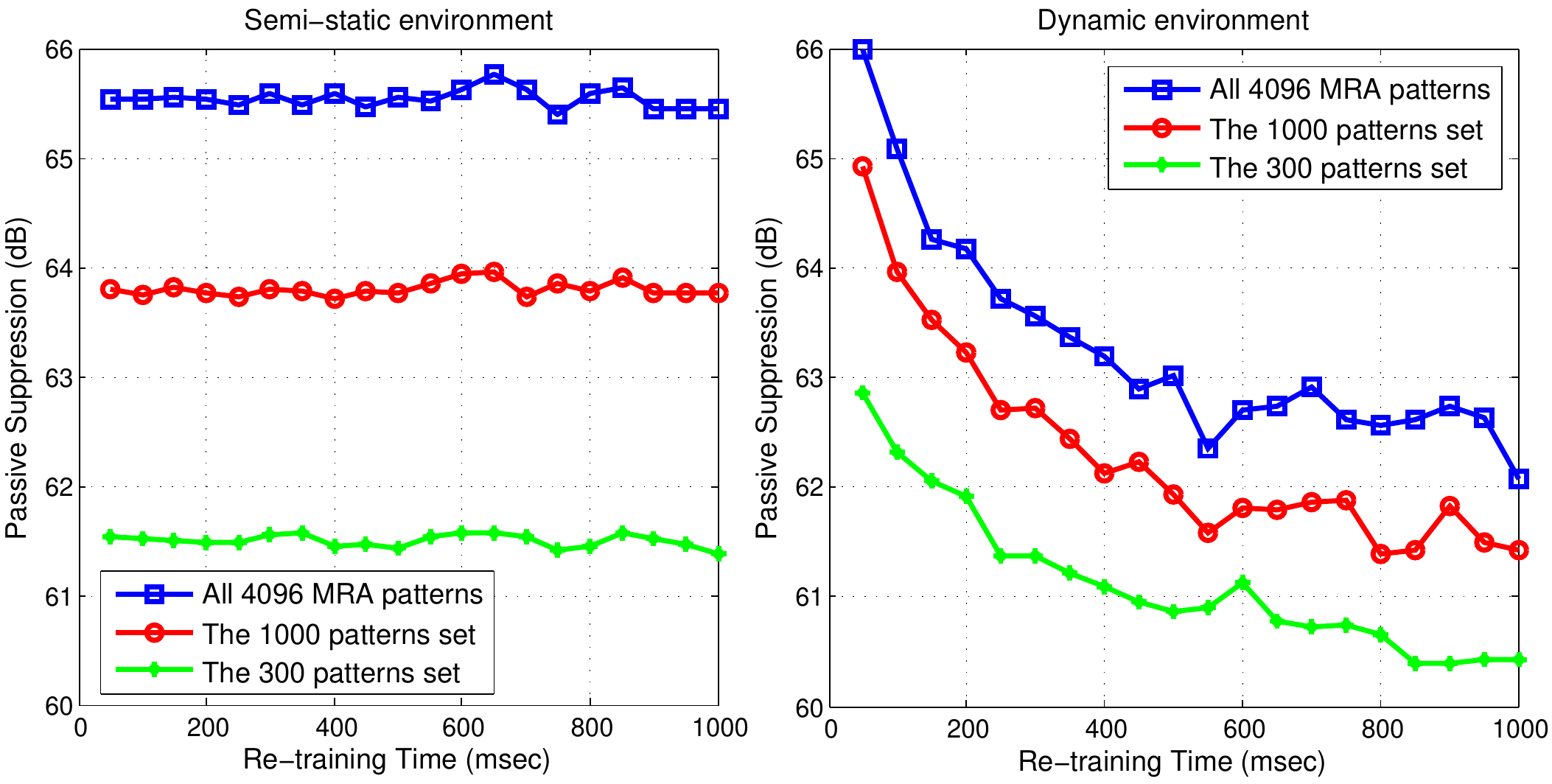}
  \caption{Passive self-interference suppression at different MRA re-training time.\label{Fig10Label}}
\end{center}
\end{figure}

In this analysis, we characterize the MRA training overhead in different environmental conditions. The system performance is evaluated for the selected suboptimal sets as well as the optimal 4096-pattern set. In this analysis, experiments are conducted in two main environments: semi-static environment and dynamic environment. Figure~\ref{Fig10Label} shows the achieved average passive self-interference suppression at different re-training times for the semi-static and dynamic environments. The conclusions from these results are multifold: first, due to the slow channel variations in the semi-static environment, the system performance is almost constant with respect to the re-training time. In this kind of environments, the MRA could be trained once per second with no performance loss. Assuming that each pattern requires 2us training time, the training duration for the 4096-, 1000-, and 300-patterns sets are $\sim$8ms, 2ms, and 0.6ms respectively. If the MRA is trained once per second, the training overhead for the 4096-, 1000-, and 300-patterns sets will be 0.8\%, 0.2\%, and 0.06\% respectively, which is negligible overhead compared to the expected 100\% capacity gain achieved by full-duplex systems.

Second, in the dynamic environment, due to the relatively fast channel variations, the system starts to lose performance with the increase of the re-training time. The results show that, 2-3dB passive self-interference suppression loss is expected when the re-training time increases from 50ms to 500ms. However, for fair comparison of the different pattern sets, the overall training overhead should be considered. Thus, rather than focusing on the re-training time, it is desired to observe performance at a fixed training overhead. For example, if the training overhead is fixed at 1\% with a 2us pattern training interval then, the 4096-, 1000-, and 300-patterns sets should be compared at re-training times of $\sim$800ms, 200ms, and 60ms respectively. Comparing the performance of the different sets at the previous re-training times we note that all different sets achieve approximately same performance.      

Another practical aspect that should be considered when discussing re-training time is the useful data frame length. Although the performance of the optimum 4096-patterns set is best among the other sets, however, for reasonable training overhead, the required re-training time for the 4096-patterns set is very high. For instance, from the previous examples, we show that for the optimal 4096-patterns set at 1\% training overhead, re-training time of 800ms is required regardless of the useful data length transmitted within the 800ms. In other words, to guarantee a 1\% training overhead, a useful data frame length of $\sim$800ms should be transmitted between the two successive MRA training intervals. Therefore, in a multi-user networks, each user should be assigned a continuous 800ms interval for data transmission, which is relatively large interval. On the other hand, the 300-patterns set requires only 60ms re-training time. Accordingly, from a practical perspective, using smaller pattern sets alleviates the constraints on the overall network prformance.

\section{Overall Full-duplex System Performance}
In this section, we characterize the overall peformance of the full-duplex system utilizing MRA. For full system performance characterization, the MRA-based passive suppression is combined with the conventional digital self-interference cancellation technique. In the full-duplex system, the received signal in the time and frequency domains can be written as
\begin{equation}\label{eq:1}
y_n=h_n^I*\left(x_n^I+z_n^T\right)+ h_n^S*\left(x_n^S+z_n^T\right)+ z_n^R\text{,}
\end{equation}
\begin{equation}\label{eq:2}
Y_k=H_k^I\left(X_k^I+Z_k^T\right)+ H_k^S\left(X_k^S+Z_k^T\right)+ Z_k^R\text{,}
\end{equation}
where $x^I$, $x^S$ are the transmitted time domain self-interference and signal-of-interest signals, $h^I$, $h^S$ are the self-interference and signal-of-interest channels, $z^T$ represents the transmitter noise, $z^R$ represents the receiver noise, $n$ is the time index, $k$ is the subcarrier index, * denotes convolution process, and uppercase letters denotes the frequency-domain representation of the corresponding time-domain signals. The digital cancellation is performed by subtracting the term $\hat{H}_k^I X_k^I$ from the received signal in~\eqref{eq:2}. $\hat{H}^I$ is an estimate for the self-interference channel, obtained using training sequences transmitted at the beginning of each data frame~\cite{Ref6}.
\subsection{Overall Self-interference Cancellation}
In this analysis, the overall self-interference cancellation achieved using MRA-based passive suppression followed by digital cancellation (DC) is characterized. The complete system framework discussed in section III is used to characterize the overall self-interference cancellation performance as follows. At the beginning, the MRA is trained and the optimum pattern is selected. Then, a sequence of data frames are transmitted from one node and the other node remains silent. Now, the received data frame contains only the self-interference signal, and the noise associated with it. The self-interference channel is estimated at the beginning of each data frame, then digital cancellation is performed. The total self-interference suppression is calculated as the ratio between the transmit power and the residual self-interference power after digital cancellation. 

\begin{figure}[!ht]
\begin{center}
\noindent
  \includegraphics[width=3in]{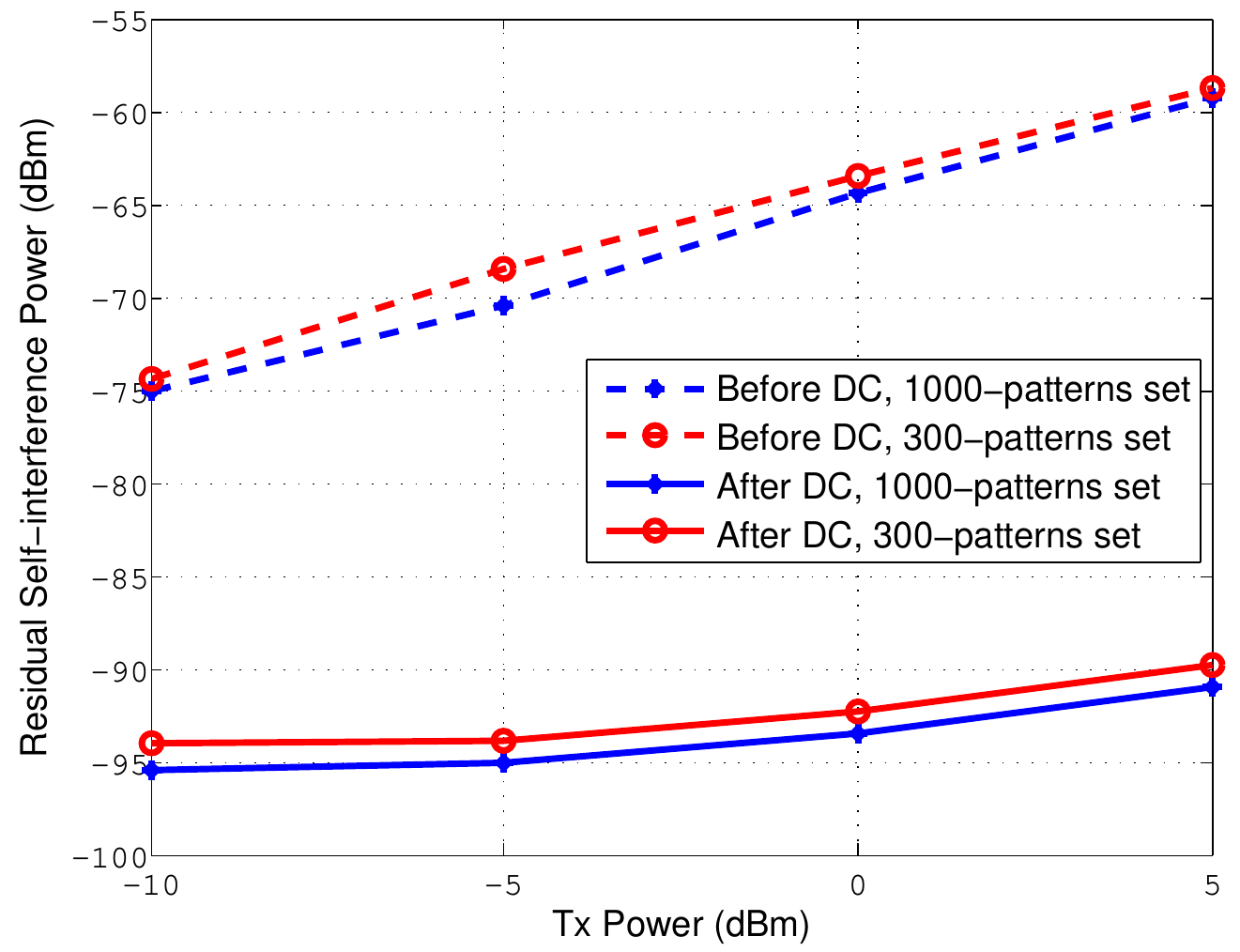}
  \caption{Residual self-interference power before and after DC at different transmit power values.\label{Fig11Label}}
\end{center}
\end{figure}

Figure~\ref{Fig11Label} shows the residual self-interference power before and after DC at different transmit power values. The results show that, in addition to the $\sim$63dB passive suppression, digital cancellation could achieve up to ~32dB more self-interference cancellation for a total of 95dB self-interference cancellation. At high transmit power values, the 32dB gain is mainly limited by the transmitter noise which can not be eliminated using conventional digital cancellation techniques. On the other hand, at low transmit power values, the achieved digital cancellation amount is limited by the receiver noise floor. At lower transmit power levels the self-interference signal is totally suppressed to below the receiver noise floor, and the full-duplex systems is expected to achieve $\sim$100\% rate gain compared to half-duplex systems. 
\subsection{Achievable Rate Gain}
The most important performance metric in full-duplex systems is the achievable rate gain compared to half-duplex systems. In this analysis, the achievable rate of the proposed full-duplex system is characterized in different experimental environments at different transmit power values. The performance is compared to the half-duplex system performance in the same environments. The achievable rate is calculated as a function of the effective Signal to Noise Ratio (SNR) as $R = log_2(1+SNR)$. One way to calculate the effective SNR in experimental analysis is by calculating the Error Vector Magnitude (EVM) defined as the distance between the received symbols (after equalization and digital cancellation) and the original transmitted symbols. Using the EVM to SNR conversion method in [26], the SNR is calculated as $SNR = 1/(EVM)^2$. The average achievable rate for both full-duplex and half-duplex systems is calculated as
\begin{equation}\label{eq:3}
R^{FD}=\frac{1}{NMK} \sum_{n=1}^N\sum_{m=1}^M\sum_{k=1}^K log_2\left[1+SINR_{n,m,k}\right]\text{,}
\end{equation}
\begin{equation}\label{eq:4}
R^{HD}=\frac{1}{NMK} \sum_{n=1}^N\sum_{m=1}^M\sum_{k=1}^K \frac{1}{2}log_2\left[1+SNR_{n,m,k}\right]\text{,}
\end{equation}
where $R^{FD}$,  $R^{HD}$ are the average achievable rate for full-duplex and half-duplex systems, $SINR$ is the effective signal to interferer plus noise ratio in full-duplex system, $SNR$ is the effective signal to noise ratio in half-duplex system, $N,M,K$ are the total number of data frames, OFDM symbols per frame, subcarriers per OFDM symbol respectively. The factor of $\frac{1}{2}$ in the half-duplex rate equation is due to the fact that each half-duplex node is transmitting only half of the time.   

\begin{figure}[!ht]
\begin{center}
\noindent
  \includegraphics[width=6.5in]{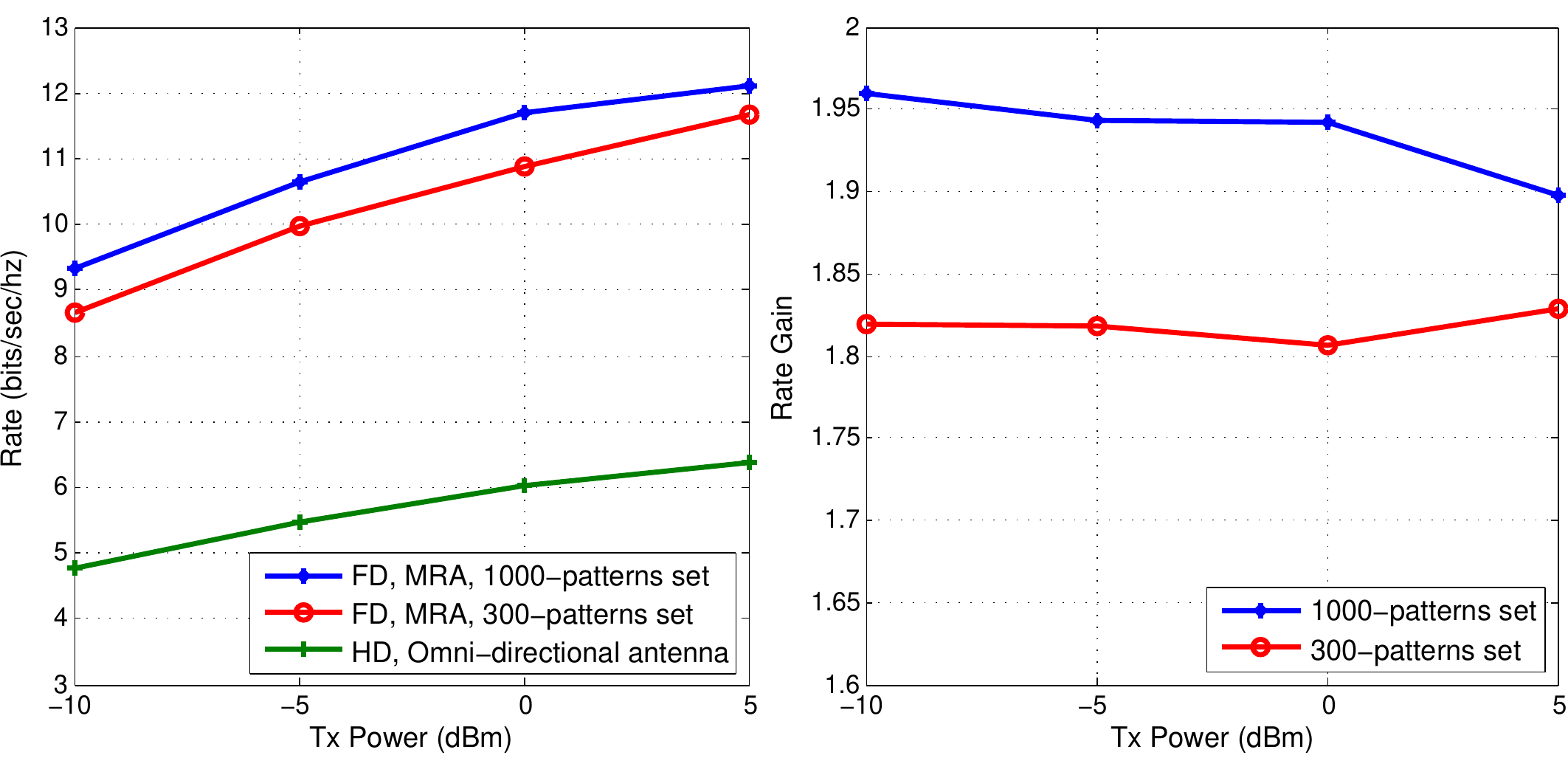}
  \caption{Overall achievable rate and rate gain for full-duplex and half-duplex systems.\label{Fig12Label}}
\end{center}
\end{figure}

Figure~\ref{Fig12Label} shows the achievable rate and the rate gain for the full-duplex and hlaf-duplex systems at different transmit power values. The results show that, the proposed full-duplex system achieves 80$-$90\% rate gain compared to the half-duplex system at 5dBm transmit power in typical indoor environments. The reason why the proposed full-duplex system could not achieve the 100\% rate gain even at low transmits power values is due to the 1dB signal-of-interest power loss shown in figure~\ref{Fig9Label}. This signal-of-interest power loss makes the full-duplex SINR less than the half-duplex SNR by 1dB even if the self-interference signal is totally suppressed below the noise floor. On the other hand, the performance difference between the 1000-patterns and the 300-patterns sets is due to the difference in the achieved self-interference cancellation amount as shown in figure~\ref{Fig11Label}.

\section{Conclusion}
In this paper, a complete full-duplex system utilizing MRAs is proposed. MRA is a reconfigurable antenna that is capable of dynamically changing its properties according to certain input configurations. The system performance is experimentally investigated in different indoor environments. The results show that, a total of 95dB self-interference cancellation is achieved by combining the MRA-based passive suppression technique with the conventional digital self-interference cancellation technique. In addition, the full-duplex achievable rate is experimentally investigated in typical indoor environments showing that, the proposed full-duplex system achieves up to 90\% rate improvement compared to half-duplex systems in typical indoor environments.

%\input{Figures}

% that's all folks
\end{document}